\documentclass[doublecol]{epl2} 

\usepackage{amsmath}
\usepackage{amssymb}
\usepackage{graphicx}
\usepackage{color}

\newcommand{\ud}{\mathrm{d}}

\newcommand{\be}{\begin{equation}}
\newcommand{\ee}{\end{equation}}

\newcommand{\ie}{i.e.}

\title{Dipole oscillations of a Fermi gas in a disordered trap: damping and localization}
\shorttitle{Dipole oscillations of a Fermi gas in a disordered trap: damping and localization}

\author{L.~Pezz\`e \and B.~Hambrecht \and L.~Sanchez-Palencia\thanks{E-mail: \email{lsp@institutoptique.fr}}}
\shortauthor{L.~Pezz\`e \etal}

\institute{Laboratoire Charles Fabry de l'Institut d'Optique, CNRS and Univ.~Paris-Sud, \\
  Campus Polytechnique, RD 128, F-91127 Palaiseau cedex, France}

\abstract{We theoretically study the dipole oscillations of an ideal Fermi gas in a disordered trap.
We show that even weak disorder induces strong damping of the oscillations
and we identify a metal-insulator crossover.
For very weak disorder, we show that damping results from a 
dephasing effect related to weak random perturbations of the energy spectrum.
For increasing disorder, we show that the Fermi gas crosses over to an insulating
regime characterized by strong-damping due to the proliferation of localized states.}

\begin{document}

\maketitle

\section{Introduction}
Ultracold atoms in disordered potentials are currently attracting considerable interest.
They offer unprecedented possibilities to revisit many 
open questions on disordered quantum systems 
with accurate experimental control of relevant parameters 
and original measurement techniques.
The spectacular progress achieved in disordered Bose-Einstein condensates (BECs)
have recently lead to the first direct observation of Anderson localization of
matter-waves~\cite{billy2008,roati2008}, showing remarkable agreement with theoretical calculations~\cite{lsp2007,damski2003}.
A future challenge to ultracold gases is the production of disordered Fermi systems.
Even better than Bose gases, they would mimic 
systems of direct relevance for condensed-matter physics,
such as dirty superconductors~\cite{desoSC} and granular metals~\cite{beloborodov2007}.
Moreover, ultracold gases provide original 
insights on transport phenomena without direct counterparts in 
traditional condensed-matter physics.
For instance, ultracold atoms trapped in harmonic potentials 
may undergo dipole oscillations -- \ie\ oscillations of the center of mass (CM) --
that can be observed for tens of periods.
Since dipole oscillations are undamped in pure harmonic traps,
irrespective to the nature of particles, temperature and interactions~\cite{Kohn},
damping results from the influence of external potentials only.
Hence, dipole oscillations have been used to characterize
several properties of quantum gases \cite{Stringari1, Stringari2}.
These include bosonic quantum degeneracy~\cite{moritz2003},
inhibition of transport of Bose gases in one-dimensional (1D) optical lattices~\cite{fertig2005},
fermionic band insulators in periodic lattices~\cite{Pezze_2004,Pezze_2004bis},
localization of spin-polarized fermions in disorder-free aperiodic lattices~\cite{Scarola_2006},
and interaction-controlled transport of fermions in optical lattices~\cite{Strohmaier_2007},
just to mention a few.

\begin{figure}[!b]
\begin{center}
\includegraphics[scale=0.6]{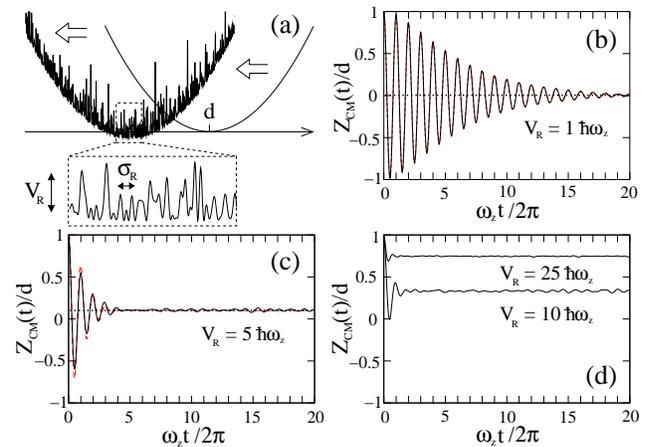}
\end{center}
\vspace{-0.5cm}
\caption{\small{
Dipole oscillations of a 3D elongated ($\lambda=8$) disordered Fermi gas.
a) The oscillations are excited by shifting the trap from $z=d$ to $z=0$
at time $t=0$.
The disorder is also switched on at $t=0$.
b-d) Dipole oscillations, averaged over 25 realizations of the disorder,
for a speckle potential with correlation length $\sigma_\textrm{R}=0.08 \, l_z$
and various amplitudes $V_{\textrm{R}}$.
The solid black lines are exact numerical calculations of $Z_\textrm{CM}(t)$
(Eqs.~(\ref{zt}),(\ref{offset})),
the dashed red lines are obtained from perturbation theory
(Eqs.~(\ref{dipole}),(\ref{gamma}),(\ref{offaverage3})).
Horizontal dotted lines are Eq.~(\ref{offset}).
Here, we have used $d=5 \,l_z$, $T=0$ and $\mu= 200 \ \hbar\omega_z$
\cite{exp_parameters}.}} 
\label{figure1}
\end{figure}

In this work, we study the dipole oscillations
of a spin-polarized Fermi gas in a harmonic trap combined
with a 1D disordered potential.
Dipole oscillations are induced by a sudden displacement of the trap center
(see Fig.~\ref{figure1}(a)), a technique which is routinely
used in experiments with ultracold atoms.
We find strong damping even for weak disorder, and identify a metal-insulator crossover
when the disorder is increased.
We show that for very weak disorder, damping results
from dephasing of different modes oscillating with frequencies randomly shifted around the 
harmonic oscillator frequency.
This leads to weakly-damped and almost centered oscillations (see Figs.~\ref{figure1}(b) and \ref{figure1}(c)),
which are accurately described by perturbation theory.
For increasing disorder, the Fermi gas crosses over to a strongly insulating regime,
characterized by strongly-damped oscillations with large offset (see Fig.~\ref{figure1}(d)).
This insulating regime is due to the proliferation of single-particle localized states.
So far, dipole oscillations in the presence of disorder have been studied only
in BECs~\cite{lye2005,albert2008}. In this case, damping is due to dissipation induced by repulsive 
interactions~\cite{albert2008}, a process completely different from the one we identify here for non-interacting Fermi gases.

\section{Framework}
We consider a Fermi gas at thermal equilibrium with temperature $T$
in a 3D axially-symmetric harmonic trap of frequencies $\omega_z$ 
in the longitudinal direction $z$ and $\omega_\perp$ in the radial directions $x$, $y$
(the aspect ratio is $\lambda=\omega_\perp/\omega_z$).
The trap is initially centered at $(x,y,z)=(0,0,d)$.
Because of the Pauli exclusion principle, the fermions populate
the single-particle eigenstates of the displaced trap,
$\Phi_{n_x,n_y,n_z}^{(d)}(x,y,z)=\phi_{n_x,n_y}(x,y)\varphi^{(d)}_{n_z}(z)$, 
according to the Fermi-Dirac distribution.
Here, $|\phi_{n_x,n_y}\rangle$ is the eigenfunction of the radial harmonic trap
associated with the eigenenergy $\hbar \omega_{\perp} (n_x+n_y+1)$. 
The function $\varphi_n^{(d)}(z) \equiv \varphi_n(z-d)$ with
$\varphi_n(z) =(n!2^n l_{z} \sqrt{\pi})^{-1/2}
\exp{(-z^2/2l_{z}^2)} H_n\left(z/l_{z}\right)$
is the eigenfunction of the 1D longitudinal harmonic potential
associated with eigenenergy $\epsilon_n = \hbar \omega_{z} (n+1/2)$,
where $H_n(z)$ is the Hermite polynomial of index $n$
and $l_{z}=\sqrt{\hbar/m\omega_{z}}$
the oscillator length in the $z$-direction.
At time $t=0$, the trap center is abruptly shifted to
$z=0$ and
a 1D homogeneous disordered potential, $V(z)=V_{\textrm{R}} v(z/\sigma_{\textrm{R}})$,
with average $\langle V \rangle=0$, 
amplitude $V_{\textrm{R}}$ 
and correlation length $\sigma_{\textrm{R}}$,
is switched on (see Fig.~\ref{figure1}(a)).
In the absence of disorder, this process induces undamped dipole oscillations
of the CM along $z$,
$Z_\textrm{CM} (t) = d \cos(\omega_z t)$~\cite{Kohn}.
In the presence of disorder, we write the CM motion,
averaged over different realizations of the disorder, as
\be \label{dipole}
Z_{\mathrm{CM}}(t) = Z_{\mathrm{osc}}(t) + Z_{\infty},
\ee
where $Z_{\mathrm{osc}}(t) \sim \Gamma (t)\cos(\omega_z t)$ is the oscillating part
with $\Gamma(t)$ an envelope function giving the {\it damping} of the oscillations
and $Z_{\infty}$ is the oscillation {\it offset}.
In the remainder of the manuscript, we evaluate the quantities $\Gamma (t)$ and $Z_{\infty}$ and
identify their physical origin.

Using the density-matrix formalism, the CM motion reads
$Z_{\mathrm{CM}}(t)=\mathrm{Tr}[\hat{z} e^{-i \hat{\mathcal{H}} t/\hbar } \, \hat{\rho}_{\mathrm{eff}}  \, e^{+i \hat{\mathcal{H}} t/\hbar }]$, 
where
\be \label{Hamiltonian}
\hat{\mathcal{H}} = -\frac{\hbar^2}{2m} \frac{\ud^2 }{\ud z^2} + \frac{1}{2} m \omega_z^2 z^2 + V(z)
\ee
is the single-particle Hamiltonian along the $z$ axis,
and $\hat{\rho}_{\mathrm{eff}}$ is an effective density matrix.
The expression of $\hat{\rho}_{\mathrm{eff}}$ is obtained by tracing out the radial
degrees of freedom, which is possible because the 3D Hamiltonian is spatially separable
and unchanged in the $x$, $y$ directions at $t=0$. We find
\be \label{rhoeff}
\hat{\rho}_{\mathrm{eff}}= \sum_n \frac{f_{\mathrm{eff}}^n(T,\mu,\lambda)}{N}
|\varphi_n^{(d)} \rangle \langle \varphi_n^{(d)} |,
\ee
where the effective Fermi distribution,
$f_{\mathrm{eff}}^n(T,\mu,\lambda) = \sum_{n_{\perp}=0}^{+\infty} \frac{n_{\perp}+1}
{e^{[\hbar\omega_z (n + \lambda n_{\tiny \perp}) -\mu]/k_\textrm{B}T} +1}$,
includes the occupation numbers of the radial-trap levels \cite{notechem} and 
$N = \sum_n f_{\mathrm{eff}}^n(T,\mu,\lambda)$
is the total number of Fermions.
In a 3D elongated Fermi gas ($\lambda=\omega_\perp/\omega_z>1$),
the 1D limit is achieved when the population of the transverse excitation modes can be neglected. This occurs
for $\hbar\omega_\perp-\mu \gg k_\textrm{B}T$
when $\mu>0$ and $\mu \gg k_\textrm{B}T$ (degenerate limit)
or
for $\hbar\omega_\perp \gg k_\textrm{B}T$
when $\mu<0$ and $|\mu| \gg k_\textrm{B}T$ (classical limit).

To evaluate explicitly $Z_{\textrm{CM}}(t)$,
we use the eigenfunctions $\{|\psi_n\rangle,\, n\in \mathbb{N}\}$ 
and the associated eigenenergies $\{E_n\}$
of Hamiltonian $\hat{\mathcal{H}}$.
The $Z_{\textrm{CM}}(t)$ can be decomposed as in Eq.~(\ref{dipole})
with
\be \label{zt}
Z_{\mathrm{osc}}(t) = \sum_{n,p; E_{n} \neq E_{p}} e^{-i (E_{n}-E_{p})t/\hbar}\, \rho_{\textrm{eff}}^{n,p} \, 
\langle \psi_{p} | \hat{z} | \psi_{n} \rangle,
\ee
and
\be \label{offset}
Z_{\infty} = \sum_{n,p; E_{n} = E_{p}} \rho_{\textrm{eff}}^{n,p} \, 
\langle \psi_{p} | \hat{z} | \psi_{n} \rangle,
\ee
where the matrix elements $\rho_{\textrm{eff}}^{n,p} \equiv  \langle \psi_{n} | \hat{\rho}_{\mathrm{eff}} | \psi_{p} \rangle$
include all the dependency on the parameters of the initial Fermi gas ($T,\mu,\lambda$) and the trap displacement $d$.
The disordered potential affects the energy spectrum
(\ie\ both $E_n$ and $|\psi_n\rangle$)
of the harmonic oscillator, which, from Eq.~(\ref{zt}), appears to have
a twofold effect on $Z_\textrm{osc}(t)$.
First, it induces random, incommensurate, energy shifts which dephase the
different oscillating components of the sum in Eq.~(\ref{zt}).
As we will see, this is the main contribution to the damping effect
for sufficiently weak disorder.
Second, it induces random modifications to the harmonic oscillator eigenfunctions,
which affect the terms $\langle \psi_{p} | \hat{z} | \psi_{n} \rangle$
and $\rho_{\textrm{eff}}^{n,p}$.
According to Eq.~(\ref{offset}), these modifications
are also responsible for the existence
of an oscillation offset, $Z_{\infty}$.

Equations~(\ref{zt}) and  (\ref{offset}) are the basis of
both our analytical and numerical calculations.
In the numerics, we use a speckle potential similar to that used in many experiments
on disordered quantum gases~\cite{billy2008,Clement_2006},
for which the auto-correlation function
reads $C(\Delta z) = \langle V(z+\Delta z) V(z)  \rangle = V_{\textrm{R}}^2 c (\Delta z / \sigma_\textrm{R})$
with $c(u)=(\sin u/u)^2$.
Figures~\ref{figure1}(b-d) show dipole oscillations of a 3D elongated 
($\lambda=8$)
Fermi gas at zero temperature for various amplitudes $V_{\textrm{R}}$ \cite{exp_parameters}.
These results show increasing damping for increasing disorder, as expected.
More precisely, we find a crossover from {\it weak-damping} (metal-like) regime for
very weak disorder (see Fig.~\ref{figure1}(b,c))
to {\it strong-damping} (insulator-like) regime characterized by a significant offset
for stronger disorder (see Fig.~\ref{figure1}(d)).
As we will show, weak-damping results from 
disorder-induced 
weak perturbations of the energy spectrum,
while strong-damping signals strong localization of
the Fermi gas.

\section{Weak-damping regime}
\begin{figure}[!t]
\begin{center}
\includegraphics[scale=0.55]{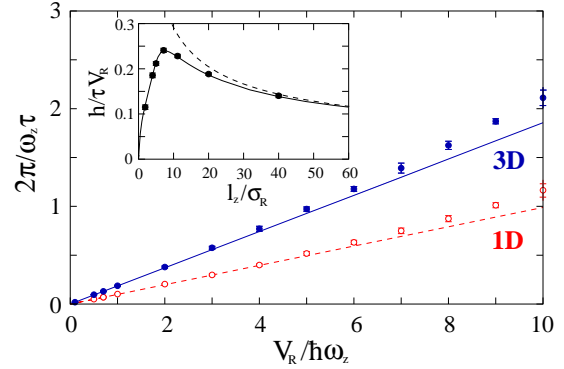}
\end{center}
\vspace{-0.5cm}
\caption{\small{Damping time, $\tau$, versus $V_{\textrm{R}}$
in 1D and 3D ($\lambda=8$) Fermi gases, 
both for $\mu=200\, \hbar\omega_z$
and $\sigma_{\textrm{R}}= 0.08~l_z$. The inset is $\tau$
as a function of $\sigma_{\textrm{R}}$ for the 3D gas.
The lines are the predictions of Eq.~(\ref{gamma}) and
the points are results of numerical simulations.
The dashed line in the inset is the white noise limit.
The parameters are $d=5~l_z$ and $T=0$.}} \label{figure2} 
\end{figure}

We first evaluate analytically $Z_{\mathrm{osc}}$
from Eq.~(\ref{zt}). We take into account first order
disorder-induced shifts to the harmonic oscillator eigenenergies and
approximate the eigenfunctions $|\psi_n\rangle$ with the harmonic oscillator ones $|\varphi_{n}^{(0)} \rangle$.
Note that, as we will show, the offset $Z_\infty$ requires second-order perturbation
calculations. For consistency in Eq.~(\ref{dipole}), one should in principle use
also second-order perturbation for calculating the envelope $Z_\textrm{osc} (t)$.
However, it leads to small corrections in $Z_\textrm{osc} (t)$ so we disregard them.
We have checked numerically that perturbation of the eigenfunctions 
has a negligible contribution to the damping, at least until almost complete dephasing has occurred.
Then, Eq.~(\ref{zt}) is significantly simplified since 
$\langle \psi_{p} | \hat{z} | \psi_{n} \rangle
\simeq \langle \varphi_{p}^{(0)} | \hat{z} | \varphi_{n}^{(0)} \rangle
= \frac{l_z}{\sqrt{2}} (\sqrt{n+1} \delta_{p, n+1} + \sqrt{n} \delta_{p, n-1})$.
Using first-order perturbation theory on the eigenenergies $E_n$ only,
we find
\be \label{Zt2}
Z_\textrm{osc} (t) = l_z \sum_{n \geq 0} \sqrt{2(n \! + \! 1)} F_{n,n+1}^{(d)}
\cos \big( \omega_z t + \delta V_n t/\hbar \big),
\ee
where 
$\delta V_n \equiv 
\langle \varphi^{(0)}_{n+1}| V(z) |\varphi^{(0)}_{n+1} \rangle -
\langle \varphi^{(0)}_n| V(z) |\varphi^{(0)}_{n} \rangle$
and
$F_{n,p}^{(d)} \equiv \langle \varphi_n^{(0)}|  \hat{\rho}_{\mathrm{eff}} | \varphi_{p}^{(0)} \rangle$.
For different realizations of the disordered potential,
$\delta V_n$ is a random quantity with
$\langle \delta V_n \rangle =0$ and 
$\langle \delta V_n^2 \rangle = V_{\textrm{R}}^2 \frac{\sigma_\textrm{R}}{l_z}
 R_n (\sigma_\textrm{R}/l_z)$, where
\be \label{sigmav}
R_n \! \Big(\frac{\sigma_\textrm{R}}{l_z}\Big) =
\int \!\! \frac{\ud \kappa}{\sqrt{\pi}}\ \tilde{c} \Big( \sqrt{2} \frac{\sigma_\textrm{R}}{l_z} \kappa \Big)
     e^{-\kappa^2} 
     \big[ L_{n+1}^{0}(\kappa^2) \! - \! L_{n}^{0}(\kappa^2) \big]^2 \nonumber
\ee
is related to the Fourier transform of the reduced correlation function of the disorder,
$\tilde{c}(q) \!\! = \!\! \int \frac{\ud u}{\sqrt{2\pi}} c(u) e^{-iqu}$, and 
$L_{n}^{\alpha}(x)$ are Laguerre polynomials~\cite{Book_Math}.
When averaging Eq.~(\ref{Zt2}) over realizations of the disorder, 
we use a Gaussian distribution in the variable $\delta V_n$ with mean square fluctuation 
$\langle \delta V_n^2 \rangle$~\cite{noteCLT}.
We then find $Z_{\mathrm{osc}}(t) \simeq d \, \Gamma(t) \cos ( \omega_z t)$, where
\be \label{gamma}
\Gamma(t) = \frac{l_z}{d} \sum_{n \geq 0}  \sqrt{2(n \! + \! 1)} F_{n,n+1}^{(d)}
\exp{\left[- \frac{V_{\textrm{R}}^2 t^2 }{2\hbar^2} \frac{\sigma_\textrm{R}}{l_z}
\mathrm{R}_n  \Big(\frac{\sigma_\textrm{R}}{l_z}\Big)\right]}.
\ee
The envelope function $\Gamma (t)$ is non-universal in the sense that
its general shape does not depend only on the disorder
(\ie\ on $V_{\textrm{R}}$, $\sigma_\textrm{R}$ and the model of disorder, $v$)
but also on the initial density matrix, \ie\
on temperature ($T$), number of fermions (or equivalently, chemical potential $\mu$),
trap geometry ($\lambda$)
and initial trap displacement ($d$).
For instance, $\Gamma (t)$ is in general neither an exponential nor a Gaussian
function~\cite{note_small_d}.
To compare Eq.~(\ref{gamma}) with numerical data, we introduce the {\it damping time} $\tau$,
defined by the equation $\Gamma(\tau) = 1/2$.
Although $\tau$ depends on both disorder and initial state of the Fermi gas,
one can obtain some universal properties from the fact that
the contribution of disorder (which appears only in the coefficient of the exponential
function in Eq.~(\ref{gamma}))
is separated from that of the initial state (in the quantities $F_{n,n+1}^{(d)}$ only).
In particular, we find $\tau \propto \hbar/|V_{\textrm{R}}|$
for any model of disorder and initial density matrix.
Numerical results for 1D and 3D Fermi gases at zero temperature
shown in Fig.~\ref{figure2} confirm this prediction.
The behavior of $\tau$ versus the correlation length $\sigma_\textrm{R}$ is more complicated.
In the white-noise limit, $\sigma_\textrm{R} \ll l_z/\sqrt{n_\textrm{F}}$,
where $n_\textrm{F}=\mu/\hbar\omega_z$
is the longitudinal Fermi level, there are many disorder peaks 
within the typical wavelength $l_z/\sqrt{n_\textrm{F}}$
of the Fermi gas.
The disordered potential felt by the wavefunctions $|\varphi_n^{(0)}\rangle$
with $n \leq n_\textrm{F}$ almost averages out.
When $\sigma_\textrm{R}$ increases, the eigenfunctions are more sensitive to the
disorder so that we expect that $\tau$ decreases.
This behavior is confirmed by Eq.~(\ref{gamma}): In the white noise limit 
the functions $R_n$ are independent of $\sigma_\textrm{R}$ so that
$\tau \propto \sqrt{l_z/\sigma_\textrm{R}}$.
The coupling of the Fermi gas to the disorder
is maximum when $\sigma_\textrm{R} \sim l_z/\sqrt{n_\textrm{F}}$~\cite{notaVn}.
Beyond, we thus expect that $\tau$ increases.
In the limit of large correlation length,
$\sigma_\textrm{R} \gg l_z \sqrt{2 n_\textrm{F}}$,
the typical size of the disorder peaks exceed the size of the Fermi gas.
Hence, the disordered potential induces only an energy shift which is approximately
equal for all the eigenstates below the Fermi energy, and
we recover harmonic oscillations (\ie\ $\tau \to \infty$).
This is again confirmed
by Eq.~(\ref{gamma}): Since the width of $\tilde{c}$ is unity
we find $R_n (\sigma_\textrm{R}/l_z) \propto (l_z/\sigma_\textrm{R})^5$
and $\tau \propto (\sigma_\textrm{R}/l_z)^2$.
These features explain the non-monotonous behavior of $\tau$ versus $\sigma_\textrm{R}$
obtained in the Inset of Fig.~\ref{figure2}.

\begin{figure}[!t]
\begin{center}
\includegraphics[scale=0.55]{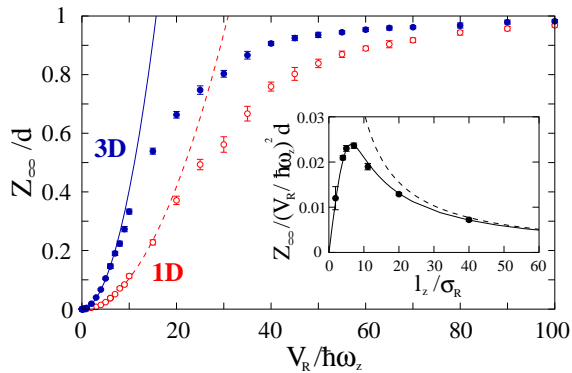}
\end{center}
\vspace{-0.5cm}
\caption{\small{Oscillation offset, $Z_\infty$, versus $V_{\textrm{R}}$
(same parameters as in Fig.~\ref{figure2}).
The points are results of numerical simulations
and the lines are Eq.~(\ref{offaverage3}).
The inset is $Z_\infty$ as a function of} $\sigma_{\textrm{R}}$ for the 3D gas.
The dashed line is the white noise approximation.
} \label{figure3} 
\end{figure}

In contrast to damping, the oscillation offset results only
from perturbation on the eigenfunctions $|\psi_n\rangle$ (see Eq.~(\ref{offset})) \cite{note_degeneracy}.
Since the average of the disorder vanishes everywhere, $\langle V(z) \rangle = 0$,
first-order perturbation is not sufficient to calculate the offset.
We thus evaluate Eq.~(\ref{offset}) up to second order, and find
\be \label{offaverage3}
Z_{\infty}  = \Big(\frac{V_{\textrm{R}}}{\hbar\omega_z}\Big)^2 \sigma_{\textrm{R}}
\sum_{\begin{subarray} 
\, \, \, n,m  \\ {n > m } \end{subarray}} \sqrt{8\frac{m!}{n!}}
\frac{F^{(d)}_{n,m}}{n-m} \, I_{n,m} \left(\!\frac{\sigma_{\textrm{R}}}{l_z}\!\right),
\ee
with
\begin{eqnarray} \label{Intnm} 
I_{n,m} \Big( \frac{\sigma_{\textrm{R}}}{l_z} \Big) = 
\int \frac{\ud \kappa}{\sqrt{\pi}} \, 
\tilde{c} \Big( \sqrt{2} \frac{\sigma_\textrm{R}}{l_z} \kappa \Big)
(i\kappa)^{n-m+1} L_{m}^{n-m}(\kappa^2) \times \nonumber \\
e^{-\kappa^2}  \big[ L_{n}^{1} (\kappa^2) - L_{m}^{1} (\kappa^2) + L_{m-1}^{1} (\kappa^2) - L_{n-1}^{1} (\kappa^2) \big] \nonumber.
\end{eqnarray}
As for damping, the behavior of the offset
as a function of $V_{\textrm{R}}$ is simple:
we find $Z_{\infty} \propto (V_{\textrm{R}}/\hbar\omega_z)^2$
for any initial state of the Fermi gas and trap displacement.
This is confirmed, for small enough values of $V_{\textrm{R}}$,
by the results of numerical calculations plotted in Fig.~\ref{figure3}.
Again, the dependence of $Z_{\infty}$ versus $\sigma_\textrm{R}$ is non-monotonous, 
as shown in the Inset of Fig.~\ref{figure3}.
In the white-noise limit, $\sigma_\textrm{R} \ll l_z/\sqrt{n_\textrm{F}}$,
we find that $Z_\infty \propto \sigma_\textrm{R}$, while for
large correlation length,
$\sigma_\textrm{R} \gg l_z \sqrt{2 n_\textrm{F}}$,
we have $Z_\infty \propto l_z^5/\sigma_{\textrm{R}}^4$.

\begin{figure*}[!ht]
\begin{center}
\includegraphics[scale=1]{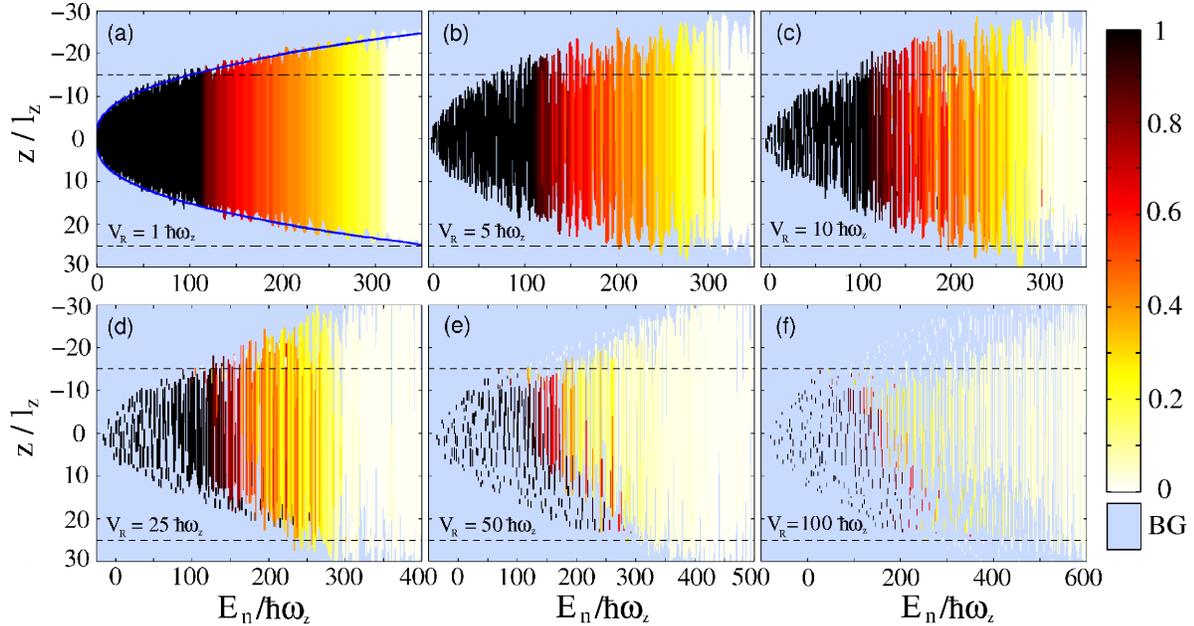}
\end{center}
\vspace{-0.5cm}
\caption{\small{Localization in a disordered trap.
Vertical lines centred in $z_n$ and of length given 
by the participation length $P_n$ are plotted at energy $E_n$. 
The color scale is given by the weight $\rho_{\textrm{eff}}^{n,n}$:
it goes from zero (white regions) to one (dark regions).
For clarity, the background (BG) is in light blue color. 
The horizontal dashed lines indicate the spatial region covered by the initial Fermi gas, i.e.
$d - l_z \sqrt{2 n_\textrm{F}} \leq z \leq d+ l_z \sqrt{2 n_\textrm{F}}$.
The different figures refer to different values of $V_\textrm{R}$ and the same realization of speckle disorder.
The solid blue line in Fig.~(a) shows the participation length $P_n^{(0)}$ 
of harmonic oscillator eigenfunctions in the absence of disorder ($V_\textrm{R}=0$).}} \label{figure4} 
\end{figure*}

Let us comment on two important properties.
First, we find that strong damping occurs even for very weak disorder.
For instance, 
$V_{\textrm{R}} = 5\hbar\omega_z$ (equal to $0.025\mu$ in Figs.~\ref{figure3} and \ref{figure4})
corresponds to $\tau \simeq 2\pi/\omega_z$, 
meaning that damping occurs typically on few
oscillations (see Fig.~\ref{figure1}(c)).
Second, we find that, for the same chemical potential,
the effect of disorder is weaker in 1D than in 3D
(the damping time $\tau$ is larger and the offset $Z_\infty$ smaller).
This can be understood intuitively from the properties of the effective Fermi distribution
$f_\textrm{eff}^n (T, \mu, \lambda)$.
Indeed, for stronger radial trapping (\ie\ for increasing $\lambda$
or in 1D compared to 3D)
and fixed Fermi energy, the population of low-energy states, which are more affected by the disorder,
decreases.
Similarly, we have found that finite temperature increases the damping time $\tau$
and decreases the offset $Z_\infty$ although the effect is quite weak.
For instance, we find that
a temperature $T=0.5~T_\textrm{F}$, where $T_\textrm{F}$ is the Fermi temperature, leads
to a small correction of about $10\%$, without affecting the general behavior of
$\tau$ and $Z_\infty$ versus the parameters.

Note also that the assumptions at the basis of our perturbation approach
are justified, {\it a posteriori}, by the excellent agreement 
between numerical calculations (points with error bars) and 
perturbation theory (lines) in Figs.~\ref{figure2} and \ref{figure3}.
This is even more striking in
Fig.~\ref{figure1}(b,c) where we plot the dipole oscillations as obtained
independently
\emph{(i)} from exact numerical calculations
(based on Eqs.~(\ref{zt}),(\ref{offset}); solid black lines)
and \emph{(ii)} from the 
prediction of Eqs.~(\ref{dipole}),(\ref{gamma}),(\ref{offaverage3})
(dashed red lines).
They are hardly distinguished in Fig.~\ref{figure1}(b,c),
indicating that perturbation theory is indeed very accurate as long as oscillations are visible
(\ie\ for $V_{\textrm{R}} \lesssim 5\hbar\omega_{z}$ for the parameters of Figs.~\ref{figure1}-\ref{figure3}).

\section{Strong-damping regime and localization} 
In the strong-damping regime (\ie\ for $V_{\textrm{R}} \gtrsim 10 \, \hbar \omega_z$
for the parameters of Figs.~\ref{figure1}-\ref{figure3}),
dipole oscillations are largely suppressed (see Fig.~\ref{figure1}(d)) and 
it is difficult to define an envelope function $\Gamma (t)$.
At the same time, the offset becomes significant (see Fig.~\ref{figure3}) and tends to $Z_{\infty} \simeq d$
for very large $V_{\textrm{R}}$ \cite{noteSr}.
This behavior signals the crossover to a strongly insulating regime
which can be related to the onset of single-particle localization.
Indeed, as shown by Eq.~(\ref{offset}), the oscillation offset $Z_\infty$ is
the sum of the average position of each eigenfunction in the disordered trap,
$z_n\equiv \langle \psi_{n} | \hat{z} | \psi_{n} \rangle$,
weighted by the corresponding population, $\rho_{\textrm{eff}}^{n,n}$~\cite{note_degeneracy}.
Extended states
--which are centered around the trap minimum ($z_n \simeq 0$)--
do not contribute to $Z_\infty$.
Conversely, localized states which are spread at random positions in the trap, $z_n \neq 0$,
may contribute to $Z_\infty$, depending on their relative population $\rho_{\textrm{eff}}^{n,n}$.

The two important features that determine the oscillation offset
are thus the localization properties of the eigenfunctions in the presence of the harmonic trap,
and their populations, which are governed by the initial, displaced Fermi gas.
In order to interpret the dipole motion in the strong-damping regime,
we now discuss these features.
The eigenfunctions $|\psi_n\rangle$ are obtained by
numerical diagonalization of Hamiltonian $\hat{\mathcal{H}}$
and characterized by two quantities:
\emph{(i)} the localization center, $z_n$, and
\emph{(ii)} the participation length, $P_n = 1/\int \ud z |\psi_n(z)|^4$,
which gives the typical extension (width) of the quantum states~\cite{notePP}.
Their populations $\rho_{\textrm{eff}}^{n,n}$ are obtained by projecting the initial
state of the Fermi gas on the eigenfunctions $|\psi_n\rangle$.
Figure~\ref{figure4} shows all these features for a 1D
Fermi gas at $T=0$, with $\mu = 200\hbar\omega$ and
initially centred at $d = 5l_z$.
For each eigenenergy $E_n$, we plot a vertical line of length $P_n$,
centred at position $z_n$, and weighted by $\rho_{\textrm{eff}}^{n,n}$ (color scale).
The different Figs.~\ref{figure4}(a)-(f) correspond to the same realization of
a speckle potential but different values of $V_\textrm{R}$.
As can be anticipated, in the weak-damping regime,
we find that the eigenfunctions are not localized (see Fig.~\ref{figure4}(a)).
The values of $P_n$ are close to the participation length $P_n^{0}$
calculated for the harmonic trap without disorder (solid blue line).
The corresponding $z_n$ slightly fluctuate around the trap center,
so that $Z_\infty \ll d$.
Increasing the amplitude $V_\textrm{R}$ of the disordered potential
(see Figs.~\ref{figure4}(b) and \ref{figure4}(c)), most of the 
populated states remain extended but become more and more perturbed.
In the crossover regime (Figs~\ref{figure4}(d) and \ref{figure4}(e)),
strongly localized states, randomly distributed in the trap, appear at low energies.
The states with lowest energy, which are more sensitive to the disorder and thus more strongly localized,
are populated when they lie in the spatial region covered by the initial Fermi gas 
(delimited by the horizontal dashed lines in Fig.~\ref{figure4}).
These states, being spatially separated from each other,
do not contribute significantly to
the sum that determines the oscillation (see Eq.~(\ref{zt})).
In contrast, they do contribute to the offset (see Eq.~(\ref{offset}))
since they are mostly located on the side of the initial gas.
The populated states with higher energy are not localized and extend
beyond the spatial region covered by the initial Fermi gas.
They can thus contribute to the oscillation, but not significantly
to the offset since their average center is close the trap center.
Finally, for very large amplitudes $V_\textrm{R}$ of the disorder (Figs.~\ref{figure4}(f)),
the Fermi gas enters the strong-damping regime.
In this case, most of the populated states are strongly localized and located in the region covered
by the initial Fermi gas, with energies about up to the chemical potential.
Note that the upper part of Figs.~\ref{figure4}(e) and \ref{figure4}(f), above the dashed line, shows 
states which are strongly localized but not populated since they lie outside the initial Fermi gas.
Some extended states with energy above the chemical potential can also, appear 
but with very small population.
The CM of the Fermi gas is thus frozen at $Z_\textrm{off}\simeq d$.
This explains qualitatively the behavior observed in Fig.~\ref{figure3}.

As we have shown, localization is a crucial ingredient for the
metal-insulator crossover investigated here.
Although localization is mainly due to the disordered potential, it is clear that
the harmonic trap significantly affects the localized states.
For instance, by virtue of finite-size effects, the lowest-energy states are localized
near the trap center (see Figs.~\ref{figure4}).
Moreover, the exponential decay of localized states,
which is the most striking signature of Anderson localization \cite{billy2008, roati2008},
is strongly suppressed since the harmonic potential dominates at large distances.
More surprisingly, our numerical data show that localized and extended states can
coexist in the same region of the spectrum at intermediate energies (see for instance Figs.~\ref{figure4}(d) and \ref{figure4}(e)),
an effect due to the trapping potential.
In the future, it will thus be interesting to study single-particle localization in disordered traps.
On one hand, it differs qualitatively form localization in homogeneous disorder,
and, on the other hand, it is directly relevant to experiments on
disordered quantum gases.

\section{Conclusions}
Dipole oscillations are an important tool for studying the dynamical properties of
ultracold gases
\cite{moritz2003,fertig2005,Pezze_2004,Pezze_2004bis,Scarola_2006,Strohmaier_2007,lye2005,albert2008,Stringari1,Stringari2}.
We have shown that dipole oscillations of trapped, disordered Fermi gases
reveal a metal-insulator crossover for increasing disorder.
Weak disorder induces weak-damping associated to weak perturbations of the energy spectrum.
Stronger disorder leads to strong-damping characterized by a large oscillation offset,
which signals the onset of localization.
We have related the insulating property of the Fermi gas to localization
of energy eigenfunctions in the disordered trap.
We have provided analytical predictions in the weak-damping regime
and numerical results in the strong-damping regime
using experimentally realistic parameters for both 1D and 3D gases.
Our predictions can thus be readily tested experimentally with
ultracold Fermi gases, in speckle potentials or other models of disorder.
Our results can be easily extended to finite temperature Fermi gases,
but we have actually found weak differences compared to the zero-temperature case.

\acknowledgments
This research was supported by the French
Centre National de la Recherche Scientifique (CNRS),
Agence Nationale de la Recherche (ANR),
Minist\`ere de l'Education Nationale, de la Recherche et de la Technologie (MENRT),
Triangle de la Physique and
Institut Francilien de Recherche sur les Atomes Froids (IFRAF).

\newpage


\begin{thebibliography}{0}
\bibitem{notechem}
Here, the zero-point energy, $\hbar\omega_z(1/2+\lambda)$,
has been included into the definition of the chemical potential $\mu$.
\bibitem{noteCLT}
We have checked numerically that this approximation is very accurate.
However complete justification of this property is beyond the scope of the present paper.
\bibitem{note_small_d}
A Gaussian envelope can be obtained in the particular case of a 1D Fermi gas at $T=0$
in the small $d$ limit.
To first order perturbation in the displacement $d$, we have 
$|\varphi_n^{(d)}\rangle \simeq |\varphi_n^{(0)}\rangle
+ \frac{d}{\sqrt{2}l_z} (\sqrt{n+1}|\varphi_{n+1}^{(0)}\rangle - \sqrt{n}|\varphi_{n-1}^{(0)}\rangle)$
and $F_{n,n+1} = \frac{d}{\sqrt{2}l_z} \frac{\sqrt{n+1}}{N} \left(f_{\mathrm{eff}}^n(T,\mu,\lambda)-f_{\mathrm{eff}}^{n+1}(T,\mu,\lambda)\right)$.
In 1D and at $T=0$, we have $N=n_\textrm{F}+1$ and
$F_{n,n+1}=\frac{d}{l_z}\frac{1}{\sqrt{2(n_\textrm{F}+1})} \delta_{n,n_\textrm{F}}$,
where $n_\textrm{F}$ is the Fermi level.
In this case only, oscillations are undamped for each realization of
the disorder with a frequency shift $\delta V_{n_\textrm{F}}$.
Then, damping results from averaging over disorder and
$\Gamma(t)=\exp{\left[- \frac{V_{\textrm{R}}^2 t^2 }{2\hbar^2} \frac{\sigma_\textrm{R}}{l_z}
\mathrm{R}_{n_\textrm{F}}  \Big(\frac{\sigma_\textrm{R}}{l_z}\Big)\right]}$
is a simple Gaussian.
Note however that this limiting case is valid only for very small displacement,
\ie\ $d \ll l_z/\sqrt{n_\textrm{F}}$.
\bibitem{notaVn} Analogous non-monotonic behavior can be found when 
evaluating the parameter $\delta V_n$ as a function of $\sigma_{\textrm{R}}$.
\bibitem{noteSr}
For the numerical calculation of this section we consider the condition of 
optimal coupling between the Fermi gas and the disordered potential, 
$\sigma_\textrm{R} \sim l_z / \sqrt{n_\textrm{F}}$.
\end{thebibliography}
\end{document}